\documentclass[12pt]{article}
\usepackage[utf8]{inputenc}

\title{ Benchmark Instances and Optimal Solutions for the {Traveling Salesman Problem with Drone}}%Flying Sidekick Traveling Salesman Problem}
\author{M. Dell'Amico$\dag$\footnote{Corresponding author}, R. Montemanni$\dag$, S. Novellani$\dag$}
\date{\small $\dag$ DISMI, University of Modena and Reggio Emilia}

\usepackage{natbib}
\usepackage{graphicx}
\usepackage{xcolor}
\usepackage{booktabs}
\usepackage{amsmath}
\usepackage{tikz}
\usepackage{hyperref}
\usepackage[normalem]{ulem}
\usetikzlibrary{arrows.meta}

\def\tr#1#2#3{\langle#1,#2,#3\rangle}

\begin{document}

\maketitle
\begin{abstract}
{The use of drones in logistics is gaining more and more interest, and {drones are becoming a more viable and common way of distributing parcels in an urban environment}. As a consequence, there is a flourishing production of articles in the field of operational optimization of the combined use of trucks and drones for fulfilling customers requests. The aim is minimizing the total time required to service all the customers, since this has obvious economical impacts. However in the literature there is not yet a widely recognized basic model, and there are not well assessed sets of instances and optimal solutions that can be considered as a benchmark to prove the effectiveness of new solution methods.
 The aim of this paper is to fill this gap. On one side we will  clearly describe some of the most common ``components" of the truck/drone routing problems and we will define nine basic  problem settings, by combining these components.
 On the other side we will consider some of the instances used by many researchers and we will provide optimal solutions for all the problem settings previously identified. Instances and detailed solutions are th{e}n organized into benchmarks made publicly available as validation tools for future research methods.}
\end{abstract}
\textbf{Keywords}: drones, FSTSP, TSP-D, benchmarks, optimal solutions

\section{Introduction}
The use of aerial drones or unmanned aerial vehicles for civil applications started in the early 1990s and had a boost with the 2000s with applications in many areas such as agriculture, inspection of structures (like bridges), land and sea monitoring,
 and logistics. {In terms of logistics applications, areal drones are currently used, for instance, to transport small automotive parts in the production line of the Audi plant of Ingolstadt, they are used for first/last mile parcel delivery (see, e.g., Google-Alphabet Wing, Amazon Prime Air, and JD.com), and for the supply of medical goods in rural or hardly connected areas (see, e.g., DHL ``Parcelcopter 4.0'' project in Tanzania or Zipline in Rwanda) as reported by the recent publication of the Roland Berger consulting group, \cite{rolabdberger}.} %{Possibly enrich the description of the applications by a few lines.}
 The operations research community has been specifically interested to study routing problems involving both trucks and drones, thus enriching the wide domain of vehicle routing problems.
 In the classic \emph{Vehicle Routing Problems} (VRP) we are given one or more vehicles based at  depots, and a set of customers. The problem asks to define routes for the vehicles so that all the customers are visited and a given objective function is optimized.  Several additional constraints and variants may be considered, such as vehicle \emph{capacity} (CVRP), \emph{time windows}  for visiting the customers (TW-VRP), \emph{pick-up} and \emph{delivery} points for the same good (PD-VRP), \emph{heterogeneous} vehicles, etc. In such environment the vehicles to be routed are trucks. When drones enter the problem we have two kinds of vehicles, namely the truck(s) and the drone(s), and both are used to serve the customers. Drones are semi-autonomous, i.e., they are carried by the truck, but sometimes can autonomously fly away from the truck, serve a customer, and return to meet the truck in some predefined  point.   Therefore a truck and a drone can serve customers in parallel, and the total time needed to serve all customers can be reduced, with respect to the truck only case.  Moreover the drone is not constrained to fly along the roads, and may have a commercial speed faster than that of the truck.

 Starting with the seminal paper by \cite{murray2015flying} many researcher tried  to propose mathematical models to represent the problem, within exact and heuristic solution methods. In a few years{,} hundreds of papers {addressing many variants of the problem} have been published. As usual when a new application {comes} to the attention of the scientific community, the researcher{s} tried to foresee the possible real life problems, and to model them, although the practical use-cases are very limited or completely absent. The emphasis of such  papers is on the possible applications, so the many proposals of different problems are not always described with the same rigour and precision.
 In this expansive and effervescent environment it is not rare to find  new variants of the problem that are presented in a paper and no more considered by the next researchers.

 A couple of recent surveys, \cite{ottooptimization} and \cite{macrina2020drone}, {try} to resume the research contributions and to give a systematic classification of problems and approaches. However, there is not yet a widely recognized  basic model, and there are not well assessed sets of instances and optimal solutions that can be considered as a benchmark to prove the effectiveness of new solution methods.
 The aim of this paper is to fill this gap. {On} one side we will  clearly describe some of the most common ``components" of the truck/drone routing problems and we will define nine basic  problem settings, by combining these components.
 {On} the other side we will consider some of the instances used by many researchers and we will provide optimal solution for all the problem settings. Instances and detailed solutions are th{e}n organized into benchmarks  made publicly available for future research.

The rest of the paper is organized as follows. In Section \ref{sec:literature} we present a brief resume of some relevant papers from the literature. In Section \ref{sec:probdesc} we formally introduce the problem and its specific components, and we give a straight mathematical formulation for each one. In Section  \ref{sec:probset} we introduce the specific problem settings that we will consider  and solve to the prove{n} optimum.
In the next  Section \ref{sec:benchmark} we  describe in detail two set of instances we propose as benchmark and how we obtained the corresponding optimal solutions for all problem settings.  %Section \ref{sec:conclusions} concludes the paper.
\section{{A b}rief literature review}\label{sec:literature}
The already cited surveys by \cite{ottooptimization} and by \cite{macrina2020drone} report more than 400 papers related to drone or truck and drone logistic applications. We are interested {into logistics problems in which} both trucks and drones contribute to the delivery. Two subclasses may be identified: (a)  the problems where the drones {and the trucks work in parallel}, i.e., the drones' {flights} serve several customers by starting and returning to the depot, while the trucks  serve other customers, and (b) the problems where the {drones are carried by the trucks and starts/returns from/to the trucks}.

The {most relevant problem of the} first {class} is {called} \emph{Parallel Drone scheduling Traveling Salesman Problem}{, presented} in  \cite{murray2015flying}, where the authors propose
a {Mixed Integer Linear Programming} (MILP) formulation  and simple greedy heuristics. {A two step iterative heuristic based on dynamic programming is proposed in \cite{mbiadou2018iterative}} {and a} mathheuristic approach is proposed in \cite{DNM2020MathH}, where mathematical models for solving some subproblems are embedded into a local search algorithm.

The second class of problems, and in particular the variation {with} a single truck and a single drone, is the subject of this paper.  This problem is called either
\emph{Flying Sidekick Traveling Salesman Problem} (FSTSP), as in \cite{murray2015flying}, or {\em TSP with Drone} (TSP-D), as in \cite{agatz2018optimization}. In the following we will refer to our problem as FSTSP, adopting the name given in the oldest paper.

All these problems are  a generalization of the \emph{Traveling Salesman Problem}, and thus are NP-hard.

In the FSTSP the drone is launched from and return{s} to the truck when it is stationed at a customer or at the depot. The drone serves a single customer and th{e}n returns to the truck. The starting and ending node of a drone {flight} must be different.  The objective is to minimize the time of return to the depot of {the latest between the} truck and {the} drone.
In the TSP-D the customers can be visited more than once, by the truck, if it is convenient for drone launching and return.
Moreover,  it is introduced the concept of \emph{loop}, i.e., the possibility for a drone to start and return in the same node. As the FSTSP by \cite{murray2015flying}, other variants of FSTSP and TSP-D, that will be described in detail in the next section, involve a time to be spent to prepare the drone before launch and a time for retrieving it and a limited duration of the battery, which, in turn, induces a constraint on the length of the {flights}. {Note that some recent works, such as \cite{murray2020multiflying} and \cite{dell2021modeling}, also consider multiple drones but they are not subject of this work.}

\cite{murray2015flying} propose a MILP formultaion and simple greedy heuristics also for the FSTSP.
\cite{agatz2018optimization} solve the TSP-D with route first-cluster second heuristics. They first build a pure TSP solution with Concorde (\cite{applegate2006concorde}) and th{e}n partition the solution to accommodate drone flights with a heuristic and an exact procedures based on dynamic programming. \cite{bouman2018dynamic} solve the TSP-D with a multiple phase approach that  firstly enumerate{s} the truck paths, and th{e}n  add{s}  the drone flights{.}

\cite{ha2015heuristic} solve the  FSTSP
with  two heuristic algorithms: a route first-cluster second and a cluster first-route second. A MILP formulation is used to solve the cluster step.
In \cite{ha2017min}, the same authors solve a similar problem with a different objective function considering separately the lengths of the truck and drone routes and the waiting times of both vehicles.
\cite{ha2020hybrid} propose an hybrid genetic algorithm improved with local search procedures
to solve both the minimum time and the minimum cost version of the FSTSP. 
In a technical report, \cite{liu2018optimization} solves a FSTSP in which at most two customers can be visited by the truck while the drone {flights}.
The author proposes a metaheuristic that is driven by two objective functions: the minimum time and the minimum consumed energy. %The proposed instances reach up to 40 customers in size.

In his PhD thesis \cite{ponza2016optimization} tackles the FSTSP proposing a  MILP formulation based on that in \cite{murray2015flying}, but
 he does not allow the drone to wait on the ground at customers nodes to save battery.   %Instances with up to 9 customers could be solved exactly in reasonable time, while the heuristic is tested on instances with up to 200 customers.

A Variable Neighborhood Descent for the FSTSP is proposed in  \cite{de2018randomized}.  
The same authors, in \cite{defreitas2018variable},  propose a hybrid general variable neighborhood search algorithm where they  first build a TSP solution thanks to {a TSP} solver, and th{e}n assign some cu{s}tomers to the drone by a greedy and a variable neighborhood search algorithm.
\cite{yurek2018decomposition} solve the FSTSP with an algorithm based on an enumeration of  the truck routes and assignment of customers to the drone,  and the construction of the full solution by solving  a MILP where the truck route is fixed.

\cite{dell2019drone} solve two variants of the FSTSP, one where the drone can wait on the ground at customer nodes and the waiting time is not included in the drone flight and one in which the drone can wait only if hovering. They call these two versions of the problem ``wait" and ``no-wait". They propose two MILP formulations that improve the Murray and Chu's one and that correctly allocate the launch times. {Further improved mathematical models are discussed in \cite{DNM2021itor}}. 
{The same authors propose some heuristics for the FSTSP ``no-wait" version in \cite{DNM2021rrls} and \cite{DNM2021bb}, leading to} some best know solutions for instances from the literature.

The TSP-D is solved by \cite{poikonen2019branch} with  a {(heuristic)} branch-and-bound method. They considers a battery limit  and drone flights starting end ending in the same node, but do not allow the {t}ruck to visit multiple times the same node.

\cite{roberti} propose exact solution approaches for TSP-D with and without loops, battery limit, and single node visits for the truck.
\cite{schermer2020b} propose MILP formulations and branch-and-cut algorithms to solve the TSP-D in which loops are allowed, battery duration is limited, only single visits are allowed, and {the} drone can wait on the ground without consuming battery. The authors also explain how to accommodate the other features in their methods but the multiple visits. %Their best methods could solve all the problems with 9 and 14 customers but only few with 20.

\section{Problem's components and mathematical models}\label{sec:probdesc}
The truck/drone problems can be described on a digraph $G=(N,A)$, where the set $N = \{0\}\cup C\cup\{n+1\}$ is the node set, with $C$ being the set of customers to be visited and $\{0\}$ and $\{n+1\}$ two nodes associated to the same physical point, namely the depot. Given  $N_0 = \{0\}\cup C$ and $N_+=C\cup \{n+1\}$, the arc set is $A=\{(i,j): i \in N_0, j \in N_+,\ i\neq j\}$.  Each arc $(i,j)$ is associated with two non-negatives traveling times: $\tau_{ij}^T$ and $\tau_{ij}^D$, that represent the time for traveling that arc by the truck and by the drone, respectively. The two matrices  are normally different.
\subsection{Problem's components}
We now describe the elements, which we call \emph{components}, that characterize the problems. Some of them are common to all problems, while other can be selected or not, thus allowing to define several different problems. In particular, we will describe nine different problem settings. {The relevant components we identified are the following ones:}\\

\noindent\emph{C0. Objective function}\\
The time instant in which both truck and drone arrive at the final depot $\{n+1\}$, must be minimized.\\

\noindent\emph{C1. Covering}\\
Each customer in $C$ must be visited exactly once by the truck or by the drone. Set $C'\subseteq C$ identifies the customers that can be served by the drone (there may be technical or organizational  reasons that forbid a drone to serve a specific customer). No service time is explicitly considered (they are assumed to be included in the traveling times).\\

\noindent\emph{C2. Truck routing}\\
The truck starts its route from node $\{0\}$ and terminate in node $\{n+1\}$ after visiting a (sub)set of customers. The route of the truck is an elementary path.\\

\begin{figure}
    \centering
\begin{tikzpicture}[thick,shorten >=1pt,->]
  \tikzstyle{vertex}=[circle,draw,minimum size=14pt,inner sep=0pt]
 \begin{scope}[shift={(0,0)}]
  \node[vertex] (i) at (0,1) {$i$};
  \node[vertex] (l) at (1.5,1) {$l$};
  \node[vertex] (k) at (3,1) {$k$};
  \node[vertex] (r) at (4.5,1) {};
  \node[vertex] (s) at (1,2) {};
  \node[vertex] (t) at (2.5,2) {};
  %\draw[] (i) -- (k);
  \draw[dashed,-{Stealth[scale=1]}] (i) -- (s);
  \draw[dashed,-{Stealth[scale=1]}] (s) -- (k);
  \draw[dashed,-{Stealth[scale=1]}] (l) -- (t);
  \draw[dashed,-{Stealth[scale=1]}] (t) -- (r);

  \draw[-{Stealth[scale=1.5]}] (i)--(l);
  \draw[-{Stealth[scale=1.5]}] (l)--(k);
  \draw[-{Stealth[scale=1.5]}] (k)--(r);
%  \draw (Q-\from) -- (Q-\to); }
\end{scope}
 \begin{scope}[shift={(6.5,0)}]
  \node[vertex] (i) at (0,1) {$i$};
  \node[vertex] (l) at (1.5,1) {$l$};
  \node[vertex] (r) at (3,1) {};
  \node[vertex] (k) at (4.5,1) {$k$};
  \node[vertex] (s) at (1.5,2.5) {};
  \node[vertex] (t) at (2,1.9) {};
  %\draw[] (i) -- (k);
  \draw[dashed,-{Stealth[scale=1]}] (i) -- (s);
  \draw[dashed,-{Stealth[scale=1]}] (s) -- (k);
  \draw[dashed,-{Stealth[scale=1]}] (l) -- (t);
  \draw[dashed,-{Stealth[scale=1]}] (t) -- (r);

  \draw[-{Stealth[scale=1.5]}] (i)--(l);
  \draw[-{Stealth[scale=1.5]}] (l)--(r);
  \draw[-{Stealth[scale=1.5]}] (r)--(k);
%  \draw (Q-\from) -- (Q-\to); }
\end{scope}
\end{tikzpicture}

    \caption{Forbidden ``crossing" sorties:  cannot be implemented by a single drone}\label{fig:crossing}
\end{figure}
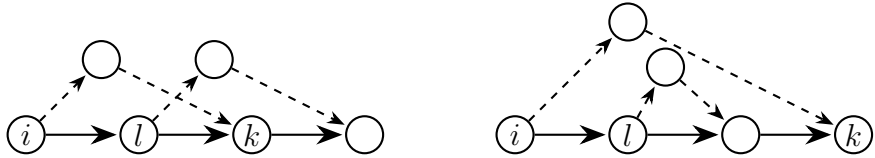

\noindent\emph{C3. Drone routing}\\
The drone is normally transported by the truck. When the truck stops in a node the drone can  fly away, serve a single customer and return to meet the truck {at}  the same node or {at} another \emph{rendezvous} node where the truck moved meanwhile. We do not allow a drone to fly away from the truck or return to it when the truck is running along an arc. Eventually the drone can start from and return to the depot, but synchronization with the truck must be guarantee (see below).
Let us call \emph{sortie} a triple $\langle i,j,k \rangle$ defining a service of the  drone that starts from $i\in N_0$, visits customer $j\in C'$, non visited by the truck, and returns to node $k\in N_+$ to meet the truck. Let $F$ be the set of all possible sorties. For each sortie $\tr{i}{j}{k}$ node $i$ must be visited by the truck before node $k$.
Since we consider problems with a single drone the sorties must not ``cross", i.e., if sortie  $\langle i,j,k \rangle$ is performed, no other sortie  $\langle l,j',k' \rangle$ can start from a node $l$, until the drone is arrived in $k$. See Figure \ref{fig:crossing} for a graphical explanation. \\

\noindent\emph{C4. Truck-drone {synchronization}}\\
 The launch of a drone and the return of the drone on the truck requires that both vehicles are present {at} the same node at the same time, i.e., a synchronization must be imposed.
 More specifically: (a) when a drone starts a {flight} from a node, the truck must be there at the same time; (b) if a drone flies from depot $\{0\}$, it must start the {flight} exactly when the truck starts its route; (c) the vehicle (drone or truck) that arrives first at a rendezvous node has to wait for the other.\\

\noindent\emph{C5. Drone loop}\\
A {flight} of a drone that starts and returns {at} the same node, when allowed,  is called a \emph{loop}. {Loops correspond to sorties of type $\langle i,j,i \rangle$.} We assume that loops from the depot are performed when the truck arrives at the depot and not when it starts from it.  \\

\noindent\emph{C6. Launch and rendezvous time}\\
When the drone starts a {flights} or returns to the truck, there are some operations that must be performed. Both truck and drone must be present at the same node to perform these operations. The time for preparing the drone at launch is given by $\sigma_L$ while rendezvous time is given by $\sigma_R$. If no launch and rendezvous time are considered we set $\sigma_L = \sigma_R = 0$.\\

\noindent\emph{C7. Launch time from the depot}\\
In some problems  the drone preparation time is not considered when the  drone flies from the starting depot 0, i.e.,  it is supposed that the launching preparation is done off-line. In this case  drone and truck do not need to spend time $\sigma_L$ before starting. When a drone loop is launched from the ending depot, $n+1$, the launch time is always considered since the truck must wait for the drone to terminate the service.\\

\noindent\emph{C8. Drone battery duration}\\
The drone has a battery limit, or \emph{endurance}, of $E$ time units, that limits the flying time. In this case we insert in set $F$ a sortie $\tr{i}{j}{k}$  only if $\tau^D_{ij}+\tau^D_{jk}+\sigma_R \le E$. {Note that in case the drone has to {hover stationary while waiting} for the truck, this waiting time has to be accounted for in the calculation of flying time.}\\

\noindent\emph{C9. Drone landing outside the truck}\\
Some problems allow the drone to land and wait for the truck, without using energy. If this landing is not allowed the drone must hover {over} the rendezvous node until the truck arrives, and this has an impact on C8, as mentioned previously. \\

Components C0 to C4 are mandatory for all problems, while C5 to C9 can be adopted or not, thus determining different problem settings. We now present a straight mathematical formulation for each of the components. Note that  we are only interested to give a clear description of the problem settings and to provide optimal benchmark solutions, so we will not address advanced (and more complex) mathematical models. For more computationally efficient models the interested reader is referred to \cite{dell2019drone}.

\subsection{Mathematical models}
Let us define the binary variables $x_{ij}$ for $(i,j) \in A$, which take value 1 if the truck travels along  arc $(i,j)$, and 0 otherwise. Moreover
for each possible drone sortie $\langle i,j,k \rangle \in F$ , we define a binary variable $y_{ijk}$,  that takes value 1 if the drone performs the sortie, and 0 otherwise.
 The non-negative variables $t_i^T, i \in N_+$ and $t_i^D, i \in N_+$ represent the time at which truck and drone, respectively, are ready to start the next activity after visiting node $i$. The time departures  from the depot are fixed at $t_0^T=t_0^D=0$. Finally a \emph{waiting} time variable $w_k, k\in N_+$ is used to quantify the time the truck waits for the arrival of the drone, at each node $k$.

 {In the following, to simplify the presentation of the mathematical models,  we initially make the hypothesis that drone loops are not allowed (i.e., set $F$ does not contain any triple of the form $\tr{i}{j}{i}$).
  Th{e}n, we remove this hypothesis and  describe the changes needed in the model to include loops. }\\

\begin{figure}[ht]
\centering
\begin{tikzpicture}[thick,shorten >=1pt,->]
  \tikzstyle{vertex}=[circle,draw,minimum size=11pt,inner sep=0pt]

  \node[vertex] (i) at (-0.2,0.65) {$i$};
  \node[vertex] (k) at (6.5,0.65) {$k$};
  \node[vertex] (j) at (1,3) {};
  \node[vertex] (j') at (4,3) {};

  %\draw[] (i) -- (k);
  \draw[dashed,-{Stealth[scale=1]}] (i) -- (j);
  \draw[dashed,-{Stealth[scale=1]}] (j') -- (k);
  \node[] (launch) at (-0.5,1.9) {$\tr{i}{\cdot}{\cdot}$};
  \node[] (return) at (6.3,1.9) {$\tr{\cdot}{\cdot}{k}$};
  %\draw[-] (0.8,0.7)--(0.8,1.3);
  %\draw[-] (4.2,0.7)--(4.2,1.3);
  %\draw[-] (5.2,0.7)--(5.2,1.3);
  \node[] (sigmaL) at (0.5,0.6) {$\sigma_L$};
  \node[] (sigmaR) at (5.5,0.6) {$\sigma_R$};
  \node[] (w) at (4.7,0.6) {$w_k$};
  \node[] (w) at (2.5,0.6) {$\tau^T_{ik}$};
  %\draw[-{Stealth[scale=1]}] (i) -- (k);
  \draw [draw=black] (0.1,0.3) rectangle (0.8,1);
  \draw [draw=black] (0.8,0.3) rectangle (4.2,1);
  \draw [draw=black] (4.2,0.3) rectangle (5.2,1);
  \draw [draw=black] (5.2,0.3) rectangle (6,1);
\end{tikzpicture}
\caption{Truck times when moving from $i$ to $k$}\label{fig:tructimes}
\end{figure}
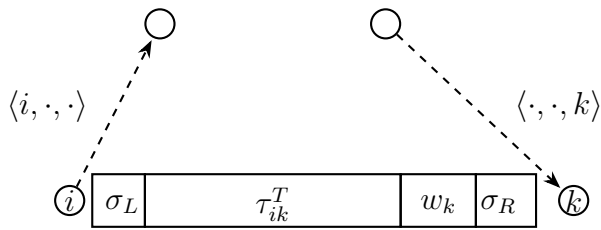
\noindent\emph{Objective function}\\
It is required to minimize the arrival at the final depot of both vehicles, i.e., $z= \max(t^T_{n+1},t^D_{n+1}\}$. It is not difficult to linearize this function and adopt it as objective, but the continuous relaxation of the corresponding model is very poor. In \cite{dell2019drone} it has been proposed an equivalent objective function which  drastically improves the bound, by  explicitly considering the contribution of the various problem elements. Let us consider  Figure \ref{fig:tructimes}, and a truck traveling directly from $i$ to $k$. Four contributions exist for the time instant in which the truck arrives  in $k$ and is ready for a next activity: (i) the traveling time $\tau^T_{ik}$; (ii) the possible launch time $\sigma_L$, if a drone sortie starts from $i$; (iii) the possible rendezvous time $\sigma_R$, if a drone sortie terminates in $k$, and (iv) the possible waiting $w_k$ for the drone arrival. The objective function is:
\begin{align}
\min z=\sum_{(i{,}j)\in A}{\tau_{ij}^T}x_{ij} + \sigma_L  \sum_{\tr{i}{r}{s}\in F} \delta^{C7}_i y_{irs} + \sigma_R \sum_{\tr{r}{s}{k}\in F} y_{rsk} +  \sum_{k\in N_+} w_k
\label{eq:obj}
\end{align}
where $\delta^{C7}_i$ is a constant set to 1 for all nodes $i\in N_+$, while $\delta^{C7}_0=1$ if component C7 (``Launch time from depot") is considered, and  $\delta^{C7}_0=0$  otherwise.\\

\noindent\emph{Covering}
\begin{align}
&\sum_{\substack{i \in N_0}} x_{ij} +  \sum_{\langle i,j,k \rangle\in F} y_{ijk} = 1 \quad  &j \in C'\label{eq:cover1}\\
&\sum_{\substack{i \in N_0}} x_{ij} = 1 \quad  &j \in C\setminus C'\label{eq:cover2}
\end{align}
Equations \eqref{eq:cover1} impose that each customer in $C'$ is served either by the truck or by the drone, while equations \eqref{eq:cover2} impose to serve the customers that can be served only by the truck. \\

\noindent\emph{Truck routing and timing}
\begin{align}
&\sum_{{j \in N_+}} x_{0j} = 1 &\label{eq:asseg1}\\%bel{eq:om1_start}\\
&\sum_{{i \in N_0}} x_{i,n+1} = 1& \label{eq:asseg2}\\%label{eq:om1_end}\\
& \sum_{{i \in N_0 }} x_{ij} = \sum_{{k \in N_+ }} x_{jk} \quad j \in C \label{eq:flow}\\
\begin{split}
& t_k^T \ge t_i^T + \tau^T_{ik} +\sigma_L\sum_{\substack{\tr{i}{r}{s}\in F}}\delta^{C7}_i  y_{irs} + \sigma_R \sum_{\tr{r}{s}{k}\in F}  y_{rsk} + w_j-M(1-x_{ik})\\
&\hspace*{0.8\textwidth} (i,k)\in A\\
\end{split}\label{eq:trucktimeA}\\
\begin{split}
& t_k^T \le t_i^T + \tau^T_{ik} +\sigma_L\sum_{\substack{\tr{i}{r}{s}\in F}}\delta^{C7}_i  y_{irs} + \sigma_R \sum_{\tr{r}{s}{k}\in F}  y_{rsk} + w_j+M(1-x_{ik})\\
&\hspace*{0.8\textwidth}(i,k)\in A\\
\end{split}\label{eq:trucktimeB}
\end{align}
As above $\delta^{C7}_i=0$ if  $i=0$ and ``Launch time from depot" is not considered, while it takes value  1 in all other cases.

Equations \eqref{eq:asseg1}-\eqref{eq:asseg2} impose that the truck exits once from depot node $\{0\}$ and terminates its route {at} depot node $\{n+1\}$, while equations \eqref{eq:flow} are the flow conservation constraints.
Timing constraints \eqref{eq:trucktimeA}-\eqref{eq:trucktimeB} compute the truck ready time in $j$ and the waiting time $w_k$ when the truck travels from $i$ to $k$ (as usual we denote with  $M$ a large positive constant).
At the same time these constraints guarantee that no subtour exists for the truck. \\

\noindent\emph{Drone routing and timing}
\begin{align}
& \sum_{{\langle i,j,k \rangle\in F}} y_{ijk}  \le \sum_{(i,h)\in A} x_{ih} &i \in N{_0} \label{eq:dronedep}\\
& \sum_{{ \langle i,j,k \rangle\in F}} y_{ijk}  \le \sum_{(h,k)\in A} x_{hk} & k \in N_+ \label{eq:dronearr}
\end{align}
Inequalities \eqref{eq:dronedep} and \eqref{eq:dronearr} impose that if a drone sortie  starts (respectfully, terminates) {at} node $i$ (resp. $k$){,} {th{e}n} the truck must exit $i$ (resp enter $k$).

The drone timing variables ares set as
\begin{align}
&t_j^D \ge t_i^T + \tau_{ij}^D +\sigma_L\delta^{C7}_i - M (1-\sum_{\tr{i}{j}{k}\in F}y_{ijk} ) \quad (i,j)\in A, j \in C' \label{eq:dronetime1}\\
&t_k^D \ge t_j^D + \tau_{jk}^D + \sigma_R - M (1-\sum_{\tr{i}{j}{k}\in F}y_{ijk} ) \quad (j,k)\in A, j \in C'\label{eq:dronetime2}
\end{align}
Crossing  sorties (see again Figure \ref{fig:crossing}) can be avoided in the following way.
Let $P(i,l)$ denote a truck path from $i \in N_0$ to $l\in C$
such that two sorties $\tr{i}{j}{k}$ and $\tr{l}{j'}{k'}$ with $k\not\in P$ exist.
In this case, the second sortie starts before the first one is returned, and the  solution is  infeasible.
Let us define $\mathcal{P}$ as the set of all the paths $P(i,l)$ with the above characteristics. {The following constraints can be derived:
\allowdisplaybreaks
\begin{align}
&  \sum_{\tr{l}{j'}{k'}\in F}y_{lj'k'} \le |P(i,l)| - \sum_{h=1}^{|P(i,l)|-1} x_{v(h)v(h+1)} - \sum_{\substack{\tr{i}{j}{k}\in F\\k\not\in P(i,l)}}y_{ijk} &P(i,l)\in \mathcal{P}  \label{eq:crossing}
\end{align}
where $P(i,l)=\{v(1),v(2),\dots, v(|P|)\}$, $v(1)=i, v(|P|)=l$.
These constraints impose that no sorties from $l$ are allowed if the truck travels along path $P(i,l)$ and a sortie starts, but does not terminate, in $P(i,l)$.

Eliminating the \emph{backward} sorties $\tr{i}{j}{k}$, where $i$ is visited by  the truck after node $k$, does not require specific constraints. Indeed, they are avoided by the truck and drone timing constraints, within the following truck-drone synchronization constraints.\\

\noindent\emph{Truck-drone synchronization}
\begin{align}
&t_0^T = 0\label{eq:tt0}\\
&t_0^D = 0\label{eq:td0}\\
&t_i^D \ge t_i^T - M (1-\sum_{(i,j)\in A} x_{ij}) \quad i \in C\label{eq:droneTruck1}\\
&t_i^D \le t_i^T + M (1-\sum_{(i,j)\in A} x_{ij}) \quad i \in C\label{eq:droneTruck2}\\
&t_k^D \ge t_k^T - M (1-\sum_{(i,k)\in A} x_{ik}) \quad k \in N_+\label{eq:droneTruck3}\\
&t_k^D \le t_k^T + M (1-\sum_{(i,k)\in A} x_{ik}) \quad k \in N_+\label{eq:droneTruck4}
\end{align}
Equations \eqref{eq:tt0} and \eqref{eq:td0} impose both drone and truck to exit the node depot $\{0\}$ at time zero. Inequalities \eqref{eq:droneTruck1}-\eqref{eq:droneTruck4} impose the synchronization of truck and drone.\\

\noindent\emph{Drone loops}\\
A {loop} is a sortie $\tr{i}{j}{i}$ starting and returning in the same node.  Allowing loops requires to insert the triples of the form $\tr{i}{j}{i}$ in $F$ and to modify  some constraints.

The drone routing constraints must allow {multiple} sorties to exit/enter in a node: a ``normal" sortie and a {a set of} loop{s}. Constraints \eqref{eq:dronedep}-\eqref{eq:dronearr} must be rewritten as  {follows:}\\
\begin{align}
& \sum_{{\langle i,j,k \rangle\in F: i\ne k}} y_{ijk}  \le \sum_{(i,h)\in A} x_{ih} &i \in N{_0} \label{eq:dronedepL1}\\
& \sum_{{ \langle i,j,k \rangle\in F: i\ne k}} y_{ijk}  \le \sum_{(h,k)\in A} x_{hk} & k \in N_+ \label{eq:dronearrL2}
\end{align}
and we must add the following constraints
\begin{align}
& \sum_{{\langle k,j,k \rangle\in F}} y_{kjk}  \le\ n \sum_{(h,k)\in A} x_{hk} &k \in N_+ \label{eq:dronedepL3}
\end{align}
Constraints \eqref{eq:dronedepL1} and \eqref{eq:dronearrL2} impose that at most one normal sortie starts from $i$ or enters $k$. Constraints \eqref{eq:dronedepL3} {allow one or more loops from a node $k$, only if the truck route enters in $k$.}

The drone-crossing constraints \eqref{eq:crossing} must be changed since many loops may start from node $l$ in a feasible solution. Therefore we must rewrite them as
\begin{align}
&  \sum_{\tr{l}{j'}{k'}\in F}y_{lj'k'} \le n(\ |P(i,l)|- \sum_{h=1}^{|P(i,l)|-1} x_{v(h)v(h+1)} - \sum_{\substack{\tr{i}{j}{k}\in F\\k\not\in P(i,l)}}y_{ijk} ) &P(i,l)\in \mathcal{P}  \label{eq:crossing2}
\end{align}

{Regarding the truck-drone synchronization constraints, we observe that when a loop $\tr{i}{j}{i}$ is performed{,} the truck must be stationary {at} node $i$ and the loop induces a pure waiting in the route. In particular, the truck waits for the time {of the flight} $\tau^D_{ij} + \tau^D_{ji}$ plus the possible launch and rendezvous times. To model this waiting{,} we can disregard the loops in the truck routing and timing constraints and add the loop time to the objective function.
Timing variables $t^T$ continue to represent a truck moving as if there are not loops, but the waiting is added to the objective function.
To this aim  we modify
\eqref{eq:trucktimeA} and \eqref{eq:trucktimeB} by removing the loops from the summations:
\begin{align}
\begin{split}
& t_k^T \ge t_i^T + \tau^T_{ik} +\sigma_L\sum_{\substack{\tr{i}{r}{s}\in F:i\neq s}}\delta^{C7}_i  y_{irs} + \sigma_R \sum_{\tr{r}{s}{k}\in F:r\neq k}  y_{rsk} + w_j\\
&\hspace*{0.5\textwidth} -M(1-x_{ik})\quad (i,k)\in A\\
\end{split}\label{eq:trucktimeA2}\\
\begin{split}
& t_k^T \le t_i^T + \tau^T_{ik} +\sigma_L\sum_{\substack{\tr{i}{r}{s}\in F:i\neq s}}\delta^{C7}_i  y_{irs} + \sigma_R \sum_{\tr{r}{s}{k}\in F:r\neq k}  y_{rsk} + w_j\\
&\hspace*{0.5\textwidth}+M(1-x_{ik})\quad (i,k)\in A\\
\end{split}\label{eq:trucktimeB2}
\end{align}
and we add the loop flying time to the objective function which becomes
\begin{align}
\begin{split}
\min z=\sum_{(i{,}j)\in A}{\tau_{ij}^T}x_{ij} + \sigma_L  \sum_{\tr{i}{r}{s}\in F} \delta^{C7}_i y_{irs} + \sigma_R \sum_{\tr{r}{s}{k}\in F} y_{rsk} \\
+  \sum_{k\in N_+} w_k + \sum_{\tr{i}{j}{i}\in F} (\tau^D_{ij} + \tau^D_{ji})y_{iji}
\end{split}
\label{eq:obj2}
\end{align}
}

Constraints \eqref{eq:dronetime1}-\eqref{eq:dronetime2} must be changed by removing the loops in the summations of the $y$ variables (i.e., writing $\sum_{\substack{\tr{i}{j}{k}\in F\\ i\ne k}}y_{ijk}$).

The battery limit constraints (see below) remain unchanged, since: (a) each time the drone returns to the truck the battery is changed, and (b) by definition, a loop $\tr{k}{l}{k}$ is  defined and inserted in set $F$ only if $\tau^D_{kl}+\tau^D_{lk}+\sigma_R \le E$.\\

\noindent\emph{Launch and rendezvous time}\\
When the launch and rendezvous times are considered{,} we set $\sigma_L$ and $\sigma_R$ at {their} proper value, otherwise we simply give them value zero.\\

\noindent\emph{Drone battery duration}\\
The endurance of the drone battery must be enough for executing a single sortie. When the drone returns on the truck it is supposed that the battery is recharged, or swapped with a fully charged one, before starting the next {sortie}.
We have thus constraints:
\begin{align}
&t_k^D -t_i^D - \sigma_L\delta^{C7}_i \le E + M (1-y_{ijk} ) \quad \tr{i}{j}{k} \in F \label{eq:endurance}
\end{align}
imposing that the sortie fly{ing} time, given by the  total sortie duration, minus the launch time, does not exceeds battery limit $E$.
Observe that for a sortie $\tr{i}{j}{k}$, the flying time includes: (i)  the time required to fly from $i$ to $j${,} and th{e}n to $k$, and (ii) the possible hovering at node $k$ to wait for the truck.
\\

\noindent\emph{Drone landing outside the truck}\\
If the drone is allowed to land at a node and wait some time {without consuming energy} in order to arrive at rendezvous node exactly with the truck, the endurance constraint \eqref{eq:endurance}
{are no longer necessary, as soon as the set $F$ of feasible drone missions is updated to only contain triplets $\tr{i}{j}{k}$ such that $\tau^D_{ij}+ \tau^D_{jk}+\sigma_R \leq E$.}\\

\section{Problem settings}\label{sec:probset}
We used the problem components described in the previous Section \ref{sec:probdesc} to obtain nine different problem settings inspired by the models presented in the literature. Since the aim of this paper is to provide clear and non-ambiguous problems descriptions and to give the corresponding optimal solutions on benchmark instances, we decided to avoid {direct citations} to problem settings already presented in the literature, {since there are cases of papers} where the text, the mathematical model, and possibly the heuristic's pseudo-codes provide different descriptions of what is supposed to be the same problem. The benchmarks here reported and the optimal solutions of our nine problem settings have the ambition to give to the researchers,  well assessed tools to validate their approaches.

Before listing our settings, we remind that components C0-C4, namely the objective function, the customer covering constraint, the topological constraints of the truck route and of the  drone sorties, and the truck-drone synchronization, are common to all problems we consider. The differences are thus given by:
drone loops, launch and rendezvous times, launch time at the depot, drone battery duration, and drone landing outside the truck.
The proposed problem settings are summarized in Table \ref{tab:settings}.

\begin{table}
\setlength{\tabcolsep}{4pt}
\begin{tabular}{lccccccccc}

                   &                   \multicolumn{9}{c}{Problem setting}\\
                   \cline{2-10}
Component                          &    1    &    2    &     3   &     4   &     5   &     6   &    7   &     8   &     9   \\
\hline
Drone loops                        &    no   &    no   &    no   &    no   &    yes  &    yes  &   yes  &    yes  &    yes  \\
Launch and rendezvous times        &   yes   &   yes   &   yes   &   yes   &    no   &    no   &   yes  &    yes  &    no   \\
Launch time at the depot           &    no   &    no   &   yes   &   yes   &    -    &     -   &   no   &    yes  &     -   \\
Drone battery duration             &   yes   &   yes   &   yes   &   yes   &   yes   &  yes    &  yes   &  yes    &  no\\
Drone landing outside the truck    &    yes  &   no    &   yes   &   no    &   yes   &  no     &  yes   &  no     &    -     \\
\hline
\end{tabular}
\caption{Benchmark problem settings}\label{tab:settings}
\end{table}

The first two problem settings are from \cite{dell2019drone}, while 3 and  4 are the variants of  1 and 2 obtained by adding the launch time for {sorties} starting at the depot. Problem settings 5 {is inspired by the} TSP-D {considered} in \cite{poikonen2019branch} and its main characteristics are: the absence of launch and rendezvous times, and to allow loops for the sorties. Problem setting 6 is a variant of setting 5 where we allow the drone to land and wait for the truck without consuming the battery. Problem settings 7 and 8 are obtained by 5 and 6 by adding the launch and rendezvous times. In problem setting 7 the launch time at the depot is considered, while in problem setting 8 it is not.
Problem setting 9 it is inspired by the work in \cite{agatz2018optimization}, but without allowing multiple visits to the same node, as done in that paper.

\section{Benchmarks}\label{sec:benchmark}
For building our benchmarks we adopted two sets of instances used by many researchers, namely the 72 instances with 10 customers given in \cite{murray2015flying} and the 100 instances with 9 customers introduced in \cite{poikonen2019branch}. We do not selected larger instances since solving these instances to the proven optimum is already challenging.

The benchmark instances and corresponding optimal solutions  are publicly available at  www.or.unimore.it  $\rightarrow$  Online Resources $\rightarrow$ Benchmark Instances and Optimal Solutions for the TSP with Drone\footnote{\href{http://www.or.unimore.it/site/home/online-resources/benchmarks-instances-and-optimal-solutions-for-the-tsp-with-drone.html}{http://www.or.unimore.it/site/home/online-resources/benchmarks-instances-and-optimal-solutions-for-the-tsp-with-drone.html}}.\\

\subsection{Instances}
\begin {sloppypar}
The two benchmarks proposed are contained in folders \textbf{DMN-B1} and \textbf{DMN-B2}. Each benchmark folder contains one subfolder for each instance, which, in turn contains 2 or 3 files describing the instance. Files \textbf{tauT.csv} and \textbf{tauD.csv} contain the time matrices $\tau^T$ and $\tau^D$. The entries have 13 decimal digits and at most two integer digits. Let $\nu$ be the number of rows and columns in one of such matrices and let $n=\nu-2$. The first and last row and column correspond to the depot nodes  $\{0\}$ and  $\{n+1\}$, respectively, while the remaining $n$ rows and columns correspond to the customer nodes. An optional file \textbf{Cprime.csv} reports, in a single row, the indices of the customers that can be served by the drone. If the file does not exist all the customers can be served by the drone.\\
\end{sloppypar}
\noindent
\textbf{Benchmark-1}\\
For our first benchmark set, \textbf{DMN-B1}, we  use the 36 well-known instances with $n=10$ randomly generated customers,  proposed by \cite{murray2015flying},  and the endurance $E$ set to either 20 or 40, thus giving 72 instances. The customers are randomly distributed across an eight-mile square region, while the depot is located in {different positions}. The nodes coordinates are given by \cite{murray2015flying} in a CSV file \textbf{nodes.csv}, but we will not include it in our benchmark-1, since it is overcome by the presence of explicit time matrices pre-computed by the authors. The truck and drone times are based, respectively,  on Manhattan distances and Euclidean  distances, and consider the truck and drone speed. The $12\times 12$ time matrices $\tau^T$ and $\tau^D$ are explicitly given in CSV files that we renamed as \textbf{tauT.csv} and \textbf{tauD.csv}. The entries of each file report a matrix with values  having 13 decimal digits. We observe that these times are not identical to the Euclidean/Manhattan distances over a truck/drone speed, but some randomization is included. We also observe that the submatrix given by the first $n+1$ rows and columns  is symmetric, while the 12-th and last row (corresponding to node $\{n+1\}$) is filled with zeros, since no arc exits the depot node $\{n+1\}$. To be precise also the first column, corresponding to arcs entering the depot node $\{0\}$, should be zero.

The  list of nodes that can be served by the drone is given in file \textbf{Cprime.csv} providing a single row with all the indices in $C'$.
The values $\sigma_L$ and $\sigma_R$ are assumed to be identical and always equal to 1.

\begin{sloppypar}
{The instances are given in separate folders with name \textbf{20140810T1234$\beta\beta$v$\xi\xi$}, where $\beta\beta$ takes one of the three values $\{37, 40, 43\}$, and $\xi\xi$ is an integer in $1,\dots, 12$.}
Each set of 12 instances refers to the same customers, but to 4 depot locations and to 3 different truck/drone speeds. \\
\end{sloppypar}

\noindent\textbf{Benchmark-2}\\
Our second benchmark set, \textbf{DMN-B2},  is based on the 100 instances with 9 customers proposed by \cite{poikonen2019branch}.
These instances   have been obtained by randomly generating $n+1$ (=10)  points in a $50\times50$ square. The first point is the depot. The original instances are given as a list of $(x,y)$ coordinates of the points.
For each arc $(i,j)\in A$ the truck travelling time is computed as the pure Manhattan  distance $\tau^{T}_{ij} = |x_i-x_j|+|y_i-y_j|$, while the drone time is given by one half the Euclidean distance, i.e. $\tau^D_{ij}=\frac{1}{2}\sqrt{(x_i-x_j)^2+(y_i-y_j)^2}$. Again we stored the time matrices in the two CSV files  \textbf{tauT.csv} and \textbf{tauD.csv} containing the $11\times 11$ matrices, with entries having 13 decimal digits. The 100 instances are contained in separate folders named {\textbf{P0}$\dots$\textbf{P99}}. In these instances all the customers can be served by the drone, so $C'=\{1,\dots,9\}$ and we do not include  a file to store this information.
The endurance is again set to 20 or 40 for all instances,  while launch and rendezvous times are set to 1 as in benchmark-1.
\subsection{Optimal solutions}\label{sec:optimal}
We implemented the  9 problem setting  of Section \ref{sec:probset} in Python and
for each of the instances of the two benchmarks we solved the mathematical models using Gurobi 9.1.1. We run our tests on an ASUS laptop with a cpu INTEL i5-7200U at 2.5Ghz, with 16 Gb of ram memory and  Windows 10 operating system. The running time ranges from few seconds up to two hours. The effectiveness of the solution method is not important for this work, so we did not tried improved models or other tricks to fasten the running time. The corresponding results are given in the files \textbf{DMN-B$x$-$ee$-solutions.csv}, where $x\in \{1,2\}$ is the benchmark number and $ee\in\{20,40\}$ is the endurance (remind that $\sigma_L$ and $\sigma_R$, when relevant, are always set to 1).
The first row of the file is a heading reporting `Instance' for the column with the instance name, `Pset$x$-opt' and `Pset$x$-sol', with $x\in\{1,\dots,9\}$, for the columns reporting the optimal solution value (13 decimal digits) and the solution of each problem setting. The solution is give{n} as a string containing the truck route and the drone {sorties}. The route is the sequence of the indices of the nodes visited by the truck, separated by blanks (with first and last index being  0 and $n+1$), while the sorties are  triplets of indices enclosed in parenthesis (where the first and last indices refers to nodes in the truck sequence, and  the mid one is the customer served by the sortie). Observe that solutions without sorties may exist as, e.g., in benchmark DMN-B2, instance P35, where the solution is an Hamiltonian tour for the truck.

\section{Conclusions}
We considered routing problems in logistics where one truck, assisted by a flying drone, has to serve a set of customers. We have listed the main `components' of such problems and combined them so as to obtain nine problem settings. We have given mathematical models for each component and setting and  used them to obtain the optimal solutions of two sets of benchmark instances. Both instances and solutions are made publicly available to foster future research in the field.

%\bibliographystyle{abbrvnat}
%\bibliography{references}
\end{document}